\documentclass{article}
\usepackage{spconf,amsmath,graphicx}

\usepackage{subcaption}
\usepackage{amsmath}
\usepackage{amssymb}
\usepackage{multirow}
\DeclareMathOperator*{\argmax}{arg\,max}

\usepackage[shortlabels]{enumitem}
\newcommand\numberthis{\addtocounter{equation}{1}\tag{\theequation}}

\title{META-LEARNING FRAMEWORK FOR END-TO-END IMPOSTER IDENTIFICATION IN UNSEEN SPEAKER RECOGNITION}
\name{Ashutosh Chaubey, Sparsh Sinha, Susmita Ghose}
\address{LG Ad Solutions}
\begin{document}
\copyrightnotice{979-8-3503-0689-7/23/\$31.00~\copyright2023 IEEE}
%
\maketitle
\begin{abstract}

Speaker identification systems are deployed in diverse environments, often different from the lab conditions on which they are trained and tested. In this paper, first, we show the problem of generalization using fixed thresholds (computed using EER metric) for imposter identification
in unseen speaker recognition and then introduce a robust speaker-specific thresholding technique for better performance. Secondly, inspired by the recent use of meta-learning techniques in speaker verification, we propose an end-to-end meta-learning framework for imposter detection which decouples the problem of imposter detection from unseen speaker identification. Thus, unlike most prior works that use some heuristics to detect imposters, the proposed network learns to detect imposters by leveraging the utterances of the enrolled speakers. Furthermore, we show the efficacy of the proposed techniques on VoxCeleb1, VCTK and the FFSVC 2022 datasets, beating the baselines by up to 10\%.
\end{abstract}
\section{Introduction}
\label{sec:intro}

Speaker recognition systems are widely used in applications related to home personalization, authentication and security. Speaker verification systems involve verifying the claimed identity of a test utterance using an utterance of the enrolled identity. Front-end speaker encoders \cite{xvector2018snyder, xvector22019snyder, zeinali2019description} first generate speaker embeddings for test and enrollment utterances. The speaker embeddings are then passed through a back-end classifier \cite{prince2007plda, GarciaRomero2011AnalysisOI}, which determines if the speakers in the two utterances are the same. Speaker identification systems involve identifying which enrolled speaker is present in the test utterance. Recently, with the advent of deep learning, several front-end speaker encoders \cite{jung2019RawNet,Desplanques2020ECAPATDNNEC,chaubey22_interspeech} and back-end classifier models \cite{heo2016bvector} have been proposed, which achieve less than 1\% equal error rates (EERs) on the VoxCeleb1 \cite{nagrani2017voxceleb1} standard trial pairs.

Speaker verification systems involve setting up a threshold for determining whether the test utterance satisfies the claimed identity or belongs to an imposter. The optimal threshold to satisfy the required false accept and reject rates vary greatly depending on the speaker characteristics and audio recording conditions (refer to Section (\ref{subsec:fixed_thres})). Imposters in speaker identification are speakers who are not enrolled on the system. The problem of imposter identification in the case of unseen speaker identification becomes challenging because of the increased confusion in the system due to multiple speakers. 

Several techniques have been proposed to increase the performance of speaker identification and verification systems in adverse conditions \cite{zhang2021tasesvnet,galina2022robust}. However, we show that the optimal performance of these techniques under different conditions is obtained at different thresholds (refer to Section (\ref{subsec:eer_thres})), making deploying such systems in unseen conditions difficult without \emph{score normalization} \cite{Reynolds2000SpeakerVUscorenorm,Matejka2017AnalysisOSscorenorm} or \emph{score calibration} \cite{mclaren2014trialbased_tbc}. 

T-Norm \cite{AUCKENTHALER200042scorenormalization} scales the similarity distribution based on the mean and variance between the test utterance and a set of cohorts. Adaptive score normalization \cite{adaptivenorm2011} involves selecting cohorts closest to the enrollment utterance for scaling. However, score normalization techniques use additional cohort utterances, which are often unavailable and lead to extra test time computation.

Saeta et al. \cite{saeta05_nolisp} and Guerra et al. \cite{guerra2008adaptive} proposed an adaptive thresholding technique using the test trial scores to improve speaker verification performance as more client scores become available. Trial-based calibration \cite{mclaren2014trialbased_tbc} converts the raw score obtained by a back-end classifier into a log-likelihood ratio using trials similar to the test trial for computing the calibration parameters. Ferrer et al. \cite{luciana2022adaptive} proposed to modify the standard back-end PLDA, introducing an adaptive calibrator which uses information such as audio duration and other related audio features to adapt to input conditions. However, these approaches could be more convenient as calibration has to be done during test time, leading to extra computation overload \cite{ferrer_reject_option}.

While currently untouched for solving the problem of imposter identification, in recent years, several meta-learning based techniques have been proposed to improve the performance of deep speaker encoders for speaker identification and verification \cite{Kumar2020DesigningNSmetalearningsurvey,wang2019protocentroid, ko2020protospeakerverification}. Prototypical networks are effective for short utterance speaker verification with an additional global classification loss to make the model more discriminative \cite{Kye2020MetaLearningFS}. Chen et al.\hspace{1mm}\cite{Chen2021Improvedmetalearningforspeakerverification} proposed additional contrastive loss and transformation coefficients for learning better speaker embeddings. Chaubey et al.\hspace{1mm}\cite{chaubey22_interspeech} proposed improved relation networks for speaker identification and an improved meta-learning training regime.

In this paper, we first demonstrate the issue with the generalizability of fixed thresholds for imposter identification and propose a robust speaker-specific thresholding technique to determine a speaker-specific threshold for identifying imposters in unseen speaker identification. In contrast to the score calibration and normalization techniques, we leverage the enrollment utterances of the speakers in the system to determine the thresholds, thus leading to a good performance without dependence on the availability of adequate test trials or cohorts.

Further, inspired by the extensive use of meta-learning for speaker identification, we propose a meta-learning approach to detect imposters in unseen speaker identification without using extra utterances during test time. The proposed imposter detection network learns the relationship the enrolled speakers have with each other and the test utterance to detect imposters. We show that the proposed network can be trained end-to-end with the speaker encoder in a meta-learning scenario. We also demonstrate that the imposter detection network during test time can detect imposters leveraging the enrollment utterances.

To show the robustness of the proposed techniques under a domain shift, we use the 2022 far-field speaker verification challenge \cite{FFSVC2022_Eval_Plan} dataset and report better performance compared to baselines. We also report the effectiveness of the proposed approach on the VCTK \cite{Veaux2016vctk} and far-field augmented VoxCeleb1 \cite{nagrani2017voxceleb1} dataset.

In summary, the following are the major contributions of this work, 
\begin{enumerate}[\itshape(i)]
    \itemsep-0.5em
    \vspace{-0.4cm}
    \item We demonstrate the issue with fixed thresholding under a domain shift and propose a simple speaker-specific thresholding technique for robust imposter identification in unseen speaker identification.
    \item We propose another novel meta-learning based imposter detection network which learns to detect imposters in unseen speaker identification.
    \item We show the efficacy of the proposed approaches in case of a domain shift on FFSVC 2022 \cite{FFSVC2022_Eval_Plan}, VCTK \cite{Veaux2016vctk} and far-field augmented VoxCeleb1 \cite{nagrani2017voxceleb1} datasets.
\end{enumerate}

The rest of this paper is organized as follows. Section (\ref{sec:speaker_id}) highlights the problem with fixed thresholding for imposter identification and introduces the proposed speaker-specific thresholding technique. Section (\ref{sec:idn}) describes the meta-learning based imposter detection network in detail. Section (\ref{sec:experiments}) details the experimental setup, and Section (\ref{sec:results}) contains the results and the ablation studies. Finally, we conclude this work in Section (\ref{sec:conclusion}). Note that we use the terms \emph{imposter detection} and \emph{imposter identification} interchangeably throughout this paper.

\section{Imposter Detection and Speaker-specific Thresholding}
\label{sec:speaker_id}

\begin{figure}
\begin{subfigure}{.5\linewidth}
  \centering
  \fbox{\includegraphics[width=0.7\linewidth,trim={2cm 2cm 4cm 1.4cm},clip]{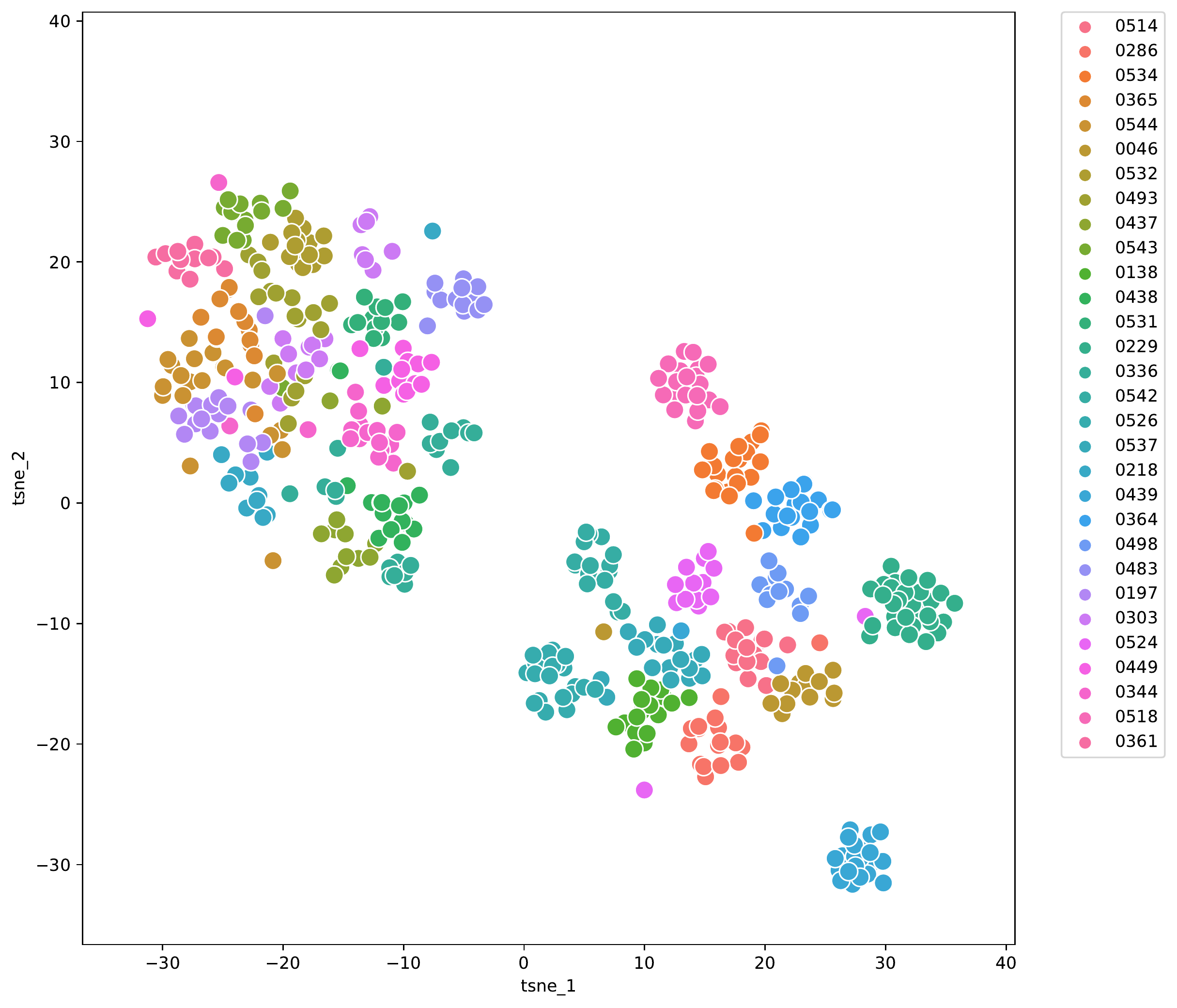}}
  \caption{VoxCeleb1}
  \label{fig:sfig1}
\end{subfigure}%
\begin{subfigure}{.5\linewidth}
  \centering
  \fbox{\includegraphics[width=0.7\linewidth,trim={2cm 2cm 4cm 1cm},clip]{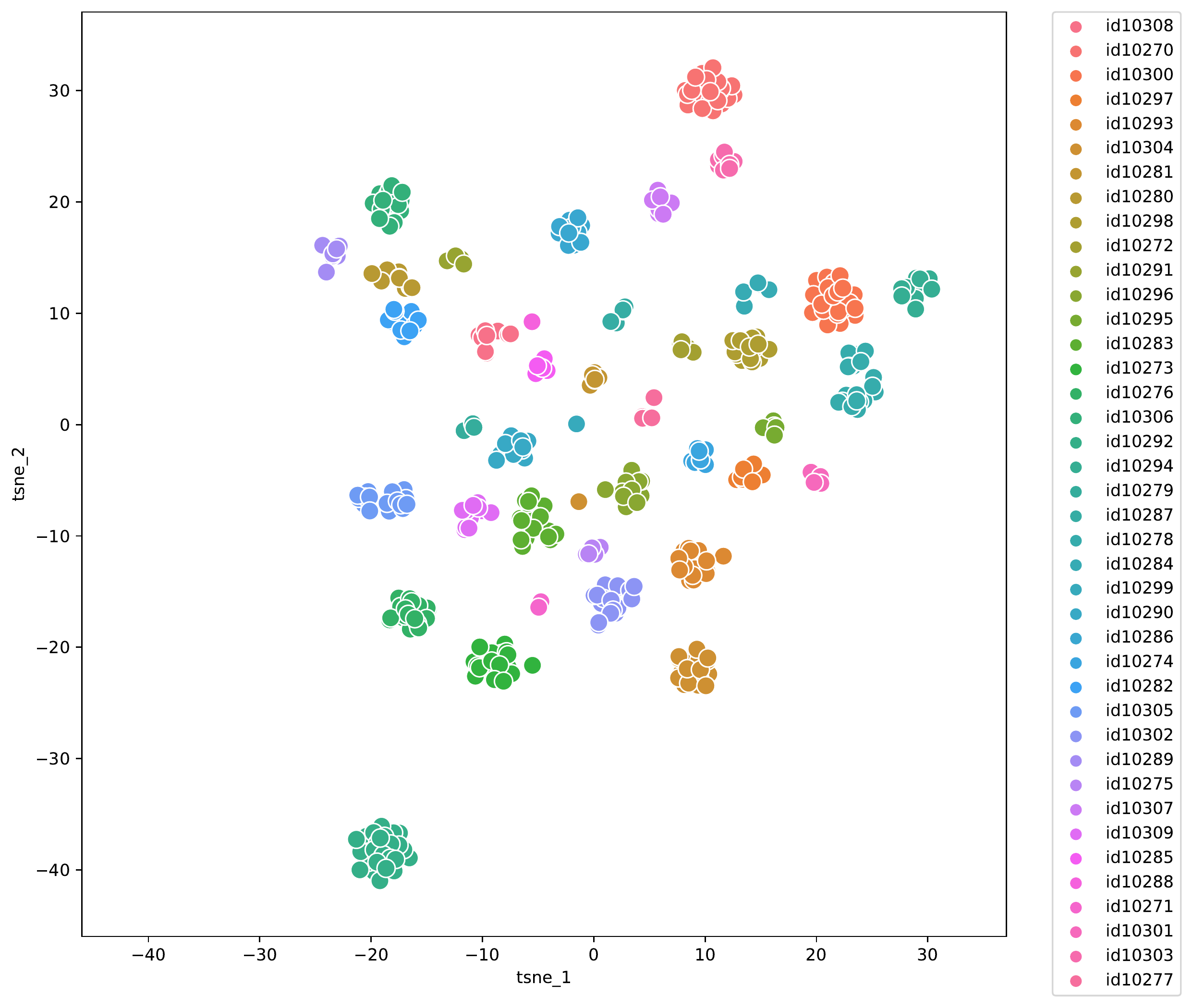}}
  \caption{SITW}
  \label{fig:sfig2}
\end{subfigure}
\caption{T-SNE plots of TDNN speaker embeddings. Speaker embeddings for SITW data are separable while those for VoxCeleb1 data are more converged and convoluted. Refer to Section \ref{sec:experiments} for data details.}
\label{fig:tsne_plot}
\end{figure}

\begin{figure*}
    \centering
    \includegraphics[width=0.7\linewidth,clip]{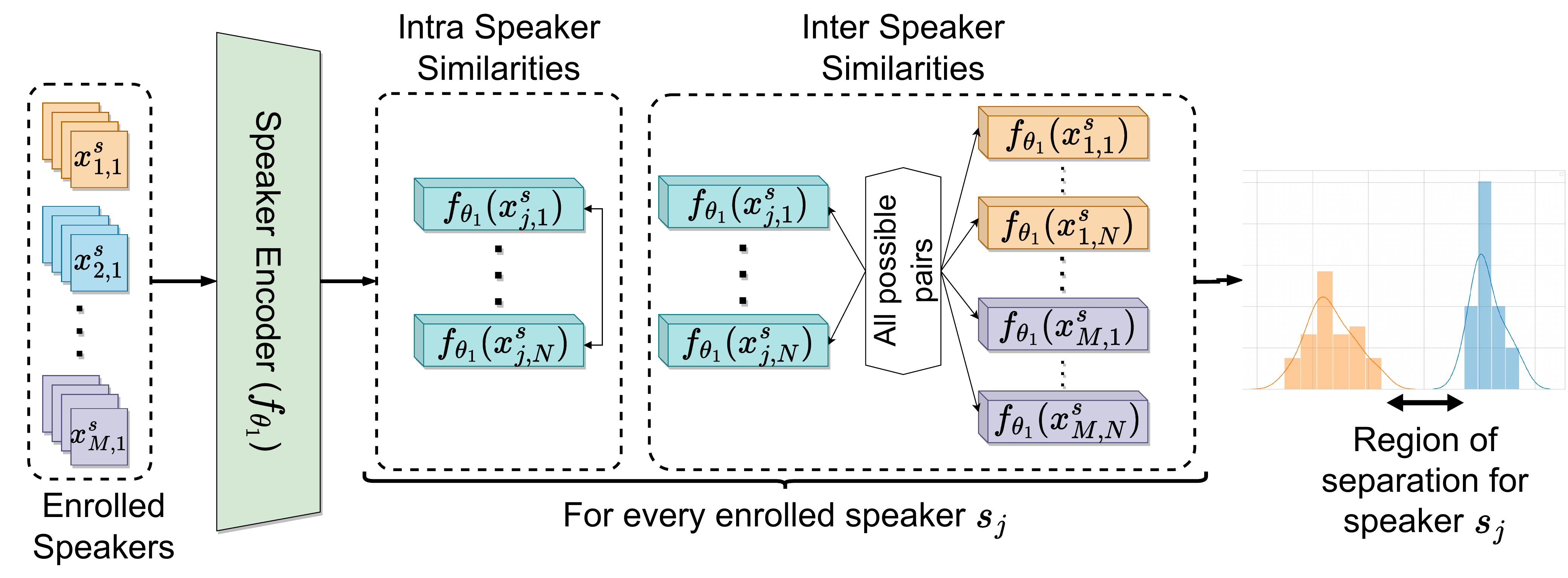}
    \caption{Speaker-specific thresholding using enrollment utterances. First the enrollment utterances are passed through the speaker encoder $f_{\theta_1}$ and then for each speaker $s_j$, threshold is computed based on inter-speaker and intra-speaker similarity distribution.}
    \label{fig:main_fig}
\end{figure*}

Speaker identification involves identifying the speaker $s_{pred}$ present in a test (query) clip $x^q_i$ from the set of already enrolled speakers ${S^{s}} = \{s_j:j \in \{1,...,M\}\}$. 
\begin{equation}
\label{eq:speaker_id}
    s_{pred} = \argmax_{s_j} \mathcal{S}(f_{\theta_1}(x_i^q), C^{s}_{j}),\text{ for } j \in \{1,...,M\}
\end{equation}
\begin{equation}
\label{eq:speaker_center}
    C^{s}_{j} = \frac{\Sigma_{k=1}^{N} f_{\theta_1}(x^{s}_{j,k})}{N}
\end{equation}

where $\mathcal{S}$ is the back-end similarity function, $M$ is the number of enrolled speakers, $N$ is the number of enrollment utterances per speaker, $x^{s}_{j,k}$ is the $k^{th}$ enrollment sample for speaker $s_j$, $C^{s}_{j}$ is the aggregated enrollment speaker embedding for speaker $s_j$, and $f_{\theta_1}$ is the speaker encoder. Unseen speaker identification constraints that the speakers present during test time are not in the training dataset.

Imposter identification or imposter detection introduces a threshold $\tau$ to this problem which forces the model to reject less confident predictions as imposters,
\begin{equation}
\label{eq:speaker_id_imposter}
    s^{*}_{pred} = \begin{cases} 
      s_{pred} & \max \mathcal{S}(f_{\theta_1}(x_i^q), C^{s}_{j}) > \tau, \\ 
      & \text{ for } j \in \{1,...,M\}\\
      imposter & otherwise 
   \end{cases}
\end{equation}

Similar to speaker verification, this threshold can be computed using the equal error rate (EER) metric or by minimizing the detection cost function (DCF). 

\subsection{Fixed thresholding}
\label{subsec:fixed_thres}
For detecting imposters, the threshold $\tau$ for a dataset $\mathcal{D}$ should be less than the intra-speaker similarities so that correct predictions are not filtered away as imposters. At the same time, the threshold should be more than inter-speaker similarities so that utterances from different speakers get filtered away. Hence, the threshold for a particular dataset $\mathcal{D}$ depends on the inter-speaker and intra-speaker separability of the utterances in the speaker embedding space, which in turn depends on the intrinsic properties of the dataset, such as speaker distribution and recording conditions. 

As shown in Figure (\ref{fig:tsne_plot}), the T-SNE \cite{vanDerMaaten2008tsne} plots for speaker embeddings corresponding to VoxCeleb1 \cite{nagrani2017voxceleb1} and SITW \cite{McLaren2016sitw} datasets have different inter-speaker separations. Clusters in Figure (\ref{fig:sfig2}) are more separable as compared to the clusters in Figure (\ref{fig:sfig1}), hence, we hypothesize that the optimal threshold for SITW would be lower compared to VoxCeleb1 data (refer to Section (\ref{subsec:eer_thres})).

\subsection{Speaker-specific thresholding}
\label{subsec:methodology}

In this work, we propose speaker-specific thresholds for imposter detection in unseen speaker identification changing Equation (\ref{eq:speaker_id_imposter}) to, 
\begin{equation}
\label{eq:speaker_id_adaptive}
    s^{**}_{pred} = \begin{cases} 
      s_{pred} & \max \mathcal{S}(f_{\theta_1}(x_i^q), C^{s}_{j}) > \tau_{s_{pred}}, \\
      & \text{ for } j \in \{1,...,M\}\\
      imposter & otherwise 
   \end{cases}
\end{equation}
where $\tau_{s_{pred}}$ is the threshold for the identified speaker. 

We leverage the enrollment utterances to compute the speaker-specific threshold for each enrolled speaker. For each enrolled speaker $s_j$, the threshold $\tau_{s_j}$ should ideally be more than the inter-speaker similarities of speaker $s_j$ with other speakers, i.e., 
\begin{align*}
\label{eq:cross_sim}
    \tau_{s_j} \geq \max \mathcal{S}(f_{\theta_1}(x^{s}_{j,k}), f_{\theta_1}(x^{s}_{u, v})),\\
    \text{ for }k, v \in \{1,...,N\} ; u \in \{1,...,M\} \text{ \& } u \neq j \numberthis
\end{align*}
where we have hypothesized that the distribution of similarity scores of $s_j$ with other enrolled speakers is a good estimation of the distribution of similarity scores of $s_j$ with any other speaker globally. Similarly, the threshold should be less than the intra-speaker similarities of speaker $s_j$, i.e.
\begin{align*}
\label{eq:auto_sim}
    \tau_{s_j} \leq \min \mathcal{S}(f_{\theta_1}(x^{s}_{j,k}), f_{\theta_1}(x^{s}_{j, l})),\\
    \text{ for }k, l \in \{1,...,N\} \numberthis
\end{align*}

Figure (\ref{fig:main_fig}) depicts the proposed speaker-specific thresholding approach. Ideally, we want the speaker-specific threshold for speaker $s_j$ to lie inside the region of separation. Due to the inefficiency of speaker encoders, the inter-speaker and intra-speaker similarity distributions are not always separable. Moreover, there are usually a few enrollment utterances available for the enrolled speakers, making the intra-speaker distribution noisy and unreliable. Thus, based on empirical results (refer to Section (\ref{subsec:imposter_id_domain_shift})), we propose the speaker-specific threshold to be, 
\begin{align*}
\label{eq:speaker_spefific_thres}
    \tau_{s_j} = \max \mathcal{S}(f_{\theta_1}(x^{s}_{j,k}), f_{\theta}(x^{s}_{u, v})),\\
    \text{ for }k, v \in \{1,...,N\} ; u \in \{1,...,M\} \text{ \& } u \neq j \numberthis
\end{align*}

\section{Meta-learning for Imposter Detection}
\label{sec:idn}

\begin{figure*}
    \centering
    \includegraphics[width=0.7\linewidth,clip]{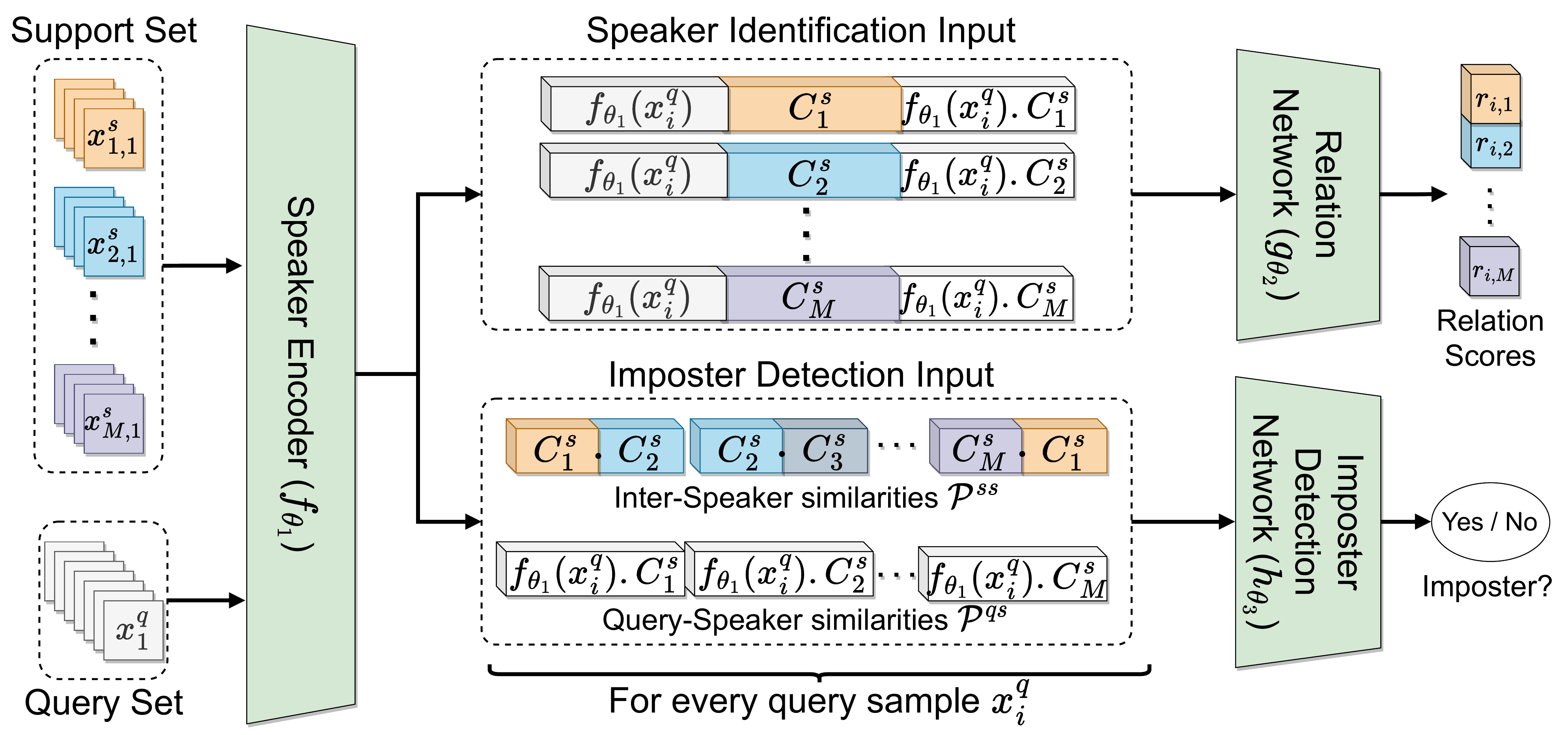}
    \caption{Meta-learning framework for end-to-end imposter identification (or imposter detection). Relation network is used for speaker identification and imposter detection network is used for detecting imposters.}
    \label{fig:main_fig2}
\end{figure*}

We leverage meta-learning for learning an imposter detection network episodically. Each episode (mini-batch) consists of a labelled \emph{support} or enrollment set containing $N$ labelled utterances for each of the $M$ enrolled speakers and an unlabelled \emph{query} or test set containing equal number of utterances from enrolled speakers and imposters. $x^s_{j, k}$ denotes the $k^{th}$ support set utterance for the $j^{th}$ enrolled speaker $s_j$. 

In this work, we detect if a query utterance $x^q_i$ belongs to an imposter or not by leveraging the enrollment utterances present in the current episode as shown in Figure (\ref{fig:main_fig2}). We decouple the problem of speaker identification from imposter detection by having two separate learnable back-ends for speaker identification and imposter detection. A relation network-based back-end proposed by Chaubey et al. \cite{chaubey22_interspeech} is used for speaker identification as it allows learning a flexible back-end and results in more discriminable speaker embeddings. For imposter identification, we propose a novel imposter detection network which uses the support set utterances along with the query utterance.

\subsection{Relation network}

First, all the utterances in the current episode (mini-batch) are encoded by the speaker encoder $f_{\theta_1}$. For each of the enrolled speakers $s_j$, we create speaker centroids $C^{s}_{j}$ by averaging out the support set embeddings similar to Equation (\ref{eq:speaker_center}). 

The relation or similarity $r_{i,j}$ between the query utterance $x^q_i$ and an enrolled speaker $s_j$ would then be, 
\begin{equation}
  r_{i, j} = g_{\theta_2}(\mathbb{C}[f_{\theta_1}(x^q_i),\text{ } C^s_j,\text{ } f_{\theta_1}(x^q_i) \cdot C^s_j])
  \label{eq:relation_bvector}
\end{equation}
where $\mathbb{C}[., ., .]$ is a channel-wise concatenation of the speaker embeddings and $g_{\theta_2}$ is the relation network. Here, we have added the additional dot product to provide an input signal related to cosine similarity (element-wise multiplication) \cite{jung2019RawNet,chaubey22_interspeech}. We train the relation network on the query utterances belonging to the enrolled (support set) speakers using the MSE objective to regress the relation or similarity score to 0 or 1,
\begin{equation}
  \mathcal{L}_{relation} = \frac{1}{M*|Q^s|}\sum_{i=1}^{|Q^s|}\sum_{j=1}^{M} (r_{i, j} - \mathfrak{1}(y^q_i == j))
  \label{eq:loss_local}
\end{equation}
where $y^q_i$ is the ground truth speaker present in the query utterance, $Q^s$ is the set of query utterances which belong to enrolled speakers and $\mathfrak{1}(.)$ yields 1 when the condition inside parenthesis is true, else 0.

\subsection{Imposter detection network}

The imposter identification network $h_{\theta_3}$ should reject all the speakers except for the enrolled speakers, and hence should learn to discriminate between the similarities that the enrolled speakers have with each other and with any of the other speakers globally (imposters). To this extent, the input to the imposter detection network is the concatenation of dot products of support set speaker centroids with each other and dot products of query embedding with speaker centroids, i.e., the imposter score $I_i$ is given by
\begin{equation}
    I_i = h_{\theta_3}(\mathbb{C}[\mathcal{P}^{ss},\text{ }\mathcal{P}^{qs}])
\end{equation}
\begin{equation}
    \mathcal{P}^{ss} = \mathbb{C}_{j=1}^{j=M}\hspace{3mm}C^s_j \cdot C^s_{(j+1)\%M}
\end{equation}
\begin{equation}
    \mathcal{P}^{qs} = \mathbb{C}_{j=1}^{j=M}\hspace{3mm}f_{\theta_1}(x_i^q) \cdot C^s_{j}
\end{equation}
where $\mathbb{C}$ is the channel-wise concatenation operator, $\mathcal{P}^{ss}$ is the input capturing inter-speaker relationships of support set speakers and $\mathcal{P}^{qs}$ is the input capturing query-support speaker relationships. Note that there can precisely be ${}^MC_2$ element-wise products in $\mathcal{P}^{ss}$ but we only take the cyclic $M$ pairs to avoid computation overload. The corresponding imposter detection loss is given by,
\begin{equation}
    \mathcal{L}_{imposter} = \frac{1}{|Q|}\sum_{i=1}^{|Q|} (I_{i} - \mathfrak{1}(x_i^q == imposter))
\end{equation}
where $Q$ is the entire query set and $\mathfrak{1}(.)$ yields 1 when the condition inside parenthesis is true, else 0.

\subsection{End-to-end training and inference}
Training the proposed framework is a multi-stage process. We first train only the speaker encoder front-end using a classification-based softmax objective, without the relation and imposter detection networks, on mini-batches without any imposter utterances in the query set. Then, we add the relation network supervising just on $\mathcal{L}_{relation}$. Finally, we train the entire framework end-to-end, on mini-batches containing both enrolled speaker and imposter utterances in the query set, i.e.
\begin{equation}
    \mathcal{L}_{total} = \mathcal{L}_{relation} + \lambda * \mathcal{L}_{imposter}
\end{equation}
Empirically, we found out that the training becomes much more stable on adding additional supervision in different stages. For all our experiments, we chose $\lambda=1$.

During inference, because the number of enrolled speakers can vary from the training time, $\mathcal{P}^{ss}$ is the concatenation of $M$ element-wise dot products of enrolled speaker pairs, and $\mathcal{P}^{qs}$ is the concatenation of $M$ element-wise dot products of query-enrolled speaker pairs. If the number of enrolled speakers is less than $M$ then the products can be repeated to get $\mathcal{P}^{ss}$ and $\mathcal{P}^{qs}$ of the required dimension as during training time, and if the enrolled speakers are more than $M$ then the products can be sampled.

\section{Experimental Setup}
\label{sec:experiments}

\subsection{Datasets and evaluation}
We use the dev set of VoxCeleb2 \cite{chung2018voxceleb2} containing 5994 speakers and over a million utterances for training the speaker encoder. As done in previous approaches \cite{Desplanques2020ECAPATDNNEC}, while training, we apply four different types of data augmentation using the Kaldi recipe with MUSAN (music, babble, noise) \cite{Snyder2015MUSANAM} and RIR (\textit{smallroom} and \textit{mediumroom} reverb simulated impulse response) \cite{ko2017rir}. 

To evaluate on the speaker verification task to compute EER thresholds, we use the standard trial pairs of VoxCeleb1 \cite{nagrani2017voxceleb1} and eval \textit{core-core} trial pairs of the SITW \cite{McLaren2016sitw} dataset. For unseen speaker identification, we run experiments on 1000 randomly generated speaker sets and report the average accuracy with 95\% confidence intervals, with each speaker having 5 enrollment clips. For computing accuracy on each speaker set, we take 10 query utterances for each of the $M$ enrolled speakers and $10M$ imposter utterances belonging to speakers not enrolled. 

We use the VoxCeleb1 test set to compute the fixed threshold leading to maximum accuracy and use that as the optimal threshold. To show the performance in case of a domain shift, we use the VCTK \cite{Veaux2016vctk} corpus and VoxCeleb1 test set with added reverberation to simulate far-field conditions. We only use the \textit{largeroom} simulated impulse response from RIR \cite{ko2017rir} dataset to augment VoxCeleb1. VCTK corpus has 109 speakers with an average of 400 utterances for each speaker, while the VoxCeleb1 test set has 40 speakers and 120 utterances per speaker on average. For performing adaptive score normalization \cite{adaptivenorm2011} within each speaker set, we randomly sample 10 cohort utterances different from the enrolled speakers within the test dataset.

Additionally, we use the 2022 FFSV Challenge \cite{FFSVC2022_Eval_Plan} development set to show how domain shift affects the performance of imposter identification. FFSVC development set has recordings from an \textit{iPhone} and an \textit{iPad} microphone, kept at \{25cm, 1m, 1.5m, 3m, 5m\} from the source. Each recording device and distance pair has recordings from 30 speakers, with an average of above 100 utterances per speaker.

\begin{table}[]
\centering
\caption{Thresholds to achieve EER for speaker verification on VoxCeleb1 \cite{nagrani2017voxceleb1} and SITW \cite{McLaren2016sitw} standard trial pairs.}
\label{tab:eer_thres}
\resizebox{\linewidth}{!}{%
\begin{tabular}{c|c|c|c} \hline \hline
\textbf{Dataset }                   & \textbf{Speaker Encoder} & \textbf{EER (\%)} & \textbf{Threshold @ EER} \\ \hline
\multirow{3}{*}{VoxCeleb1-O} & TDNN            & 0.96    & 0.311                \\
                           & Resnet50        &  2.04   &  0.516               \\
                           & RawNet3         & 0.89    &  0.304               \\ \hline
\multirow{3}{*}{SITW}      & TDNN            & 2.06    &  0.240               \\
                           & Resnet50        &  3.85   &   0.478             \\
                           & RawNet3         & 3.21    &  0.184              \\
\hline \hline
\end{tabular}
}
\end{table}

\subsection{Training configuration}
We use a TDNN-based speaker encoder \cite{Desplanques2020ECAPATDNNEC} (see Appendix for architecture) where input features are 80-dimensional mel filterbanks with a window size of 25ms and a hop of 10ms. Mean normalization is applied to the input features. SpecAugment \cite{park2019specaug} is applied on the filterbank by randomly masking 0-10 frames in the time axis and 0-8 frames in the frequency axis. Voice activity detection (VAD) is not applied anywhere. We use an Adam optimizer with $\beta = \{0.99, 0.999\}$ and with a weight decay of 2e-5. 

For training the relation and imposter detection networks, within each episode (mini-batch), we have 80 enrolled speakers, with 1 support utterance and 2 query utterances. For balanced training, there are 160 imposter utterances per episode. Relation and imposter detection networks are fully connected with ReLU activations for better gradient flow and dropout to avoid overfitting. All our experiments have been conducted on a single NVIDIA GeForce RTX 3090 GPU.

\begin{table}[]
\caption{Performance of the proposed techniques on VCTK \cite{Veaux2016vctk} and VoxCeleb1 \cite{nagrani2017voxceleb1} (with reverb) for unseen speaker identification.}
\label{tab:speaker_id}
\resizebox{\linewidth}{!}{
\begin{tabular}{c|c||cc|cc}
\hline \hline
\multirow{2}{*}{\begin{tabular}[c]{@{}c@{}}\textbf{Num.}\\ \textbf{Spkrs.}\end{tabular}} & \multirow{2}{*}{\textbf{Method}}                            & \multicolumn{2}{c|}{\textbf{VCTK}}                                                                                                               & \multicolumn{2}{c}{\begin{tabular}[c]{@{}c@{}}\textbf{VoxCeleb1}\\ \textbf{(with reverb)}\end{tabular}}                                                   \\ \cline{3-6} 
                                                                       &                                                       & \multicolumn{1}{c|}{\begin{tabular}[c]{@{}c@{}}Overall \\ (\%)\end{tabular}} & \begin{tabular}[c]{@{}c@{}}Imposter \\ (\%)\end{tabular} & \multicolumn{1}{c|}{\begin{tabular}[c]{@{}c@{}}Overall \\ (\%)\end{tabular}} & \begin{tabular}[c]{@{}c@{}}Imposter \\ (\%)\end{tabular} \\ \hline \hline
\multirow{4}{*}{M=5}                                                   & Fixed                                                 & \multicolumn{1}{c|}{95.61\begin{tiny}$\pm$0.32\end{tiny}}                                                   & 91.38\begin{tiny}$\pm$0.27\end{tiny}                                                    & \multicolumn{1}{c|}{94.64\begin{tiny}$\pm$0.35\end{tiny}}                                                   & 89.52\begin{tiny}$\pm$0.30\end{tiny}                                                    \\
                                                                       & \begin{tabular}[c]{@{}c@{}}Score \\ Norm \cite{adaptivenorm2011}\end{tabular} & \multicolumn{1}{c|}{96.58\begin{tiny}$\pm$0.19\end{tiny}}                                                   & 94.62\begin{tiny}$\pm$0.24\end{tiny}                                                    & \multicolumn{1}{c|}{94.67\begin{tiny}$\pm$0.27\end{tiny}}                                                   & 92.91\begin{tiny}$\pm$0.25\end{tiny}                                                    \\
                                                                       & SST (\begin{footnotesize}\textbf{Ours}\end{footnotesize})                                             & \multicolumn{1}{c|}{97.73\begin{tiny}$\pm$0.28\end{tiny}}                                                   & 97.38\begin{tiny}$\pm$0.31\end{tiny}                                                    & \multicolumn{1}{c|}{96.17\begin{tiny}$\pm$0.18\end{tiny}}                                                   & 94.49\begin{tiny}$\pm$0.23\end{tiny}                                                    \\
                                                                       & IDN (\begin{footnotesize}\textbf{Ours}\end{footnotesize})                                            & \multicolumn{1}{c|}{\textbf{98.06\begin{tiny}$\pm$0.26\end{tiny}}}                                          & \textbf{97.96\begin{tiny}$\pm$0.21\end{tiny}}                                           & \multicolumn{1}{c|}{\textbf{97.94\begin{tiny}$\pm$0.29\end{tiny}}}                                          & \textbf{96.82\begin{tiny}$\pm$0.22\end{tiny}}                                           \\ \hline
\multirow{4}{*}{M=10}                                                  & Fixed                                                 & \multicolumn{1}{c|}{94.82\begin{tiny}$\pm$0.26\end{tiny}}                                                   & 89.92\begin{tiny}$\pm$0.24\end{tiny}                                                    & \multicolumn{1}{c|}{93.06\begin{tiny}$\pm$0.29\end{tiny}}                                                   & 86.47\begin{tiny}$\pm$0.35\end{tiny}                                                    \\
                                                                       & \begin{tabular}[c]{@{}c@{}}Score \\ Norm \cite{adaptivenorm2011}\end{tabular} & \multicolumn{1}{c|}{96.32\begin{tiny}$\pm$0.26\end{tiny}}                                                   & \textbf{95.89\begin{tiny}$\pm$0.21\end{tiny}}                                           & \multicolumn{1}{c|}{93.78\begin{tiny}$\pm$0.32\end{tiny}}                                                   & 89.01\begin{tiny}$\pm$0.30\end{tiny}                                                    \\
                                                                       & SST (\begin{footnotesize}\textbf{Ours}\end{footnotesize})                                            & \multicolumn{1}{c|}{\textbf{96.55\begin{tiny}$\pm$0.25\end{tiny}}}                                          & 94.99\begin{tiny}$\pm$0.18\end{tiny}                                                    & \multicolumn{1}{c|}{94.28\begin{tiny}$\pm$0.29\end{tiny}}                                                   & 90.82\begin{tiny}$\pm$0.22\end{tiny}                                                    \\
                                                                       & IDN (\begin{footnotesize}\textbf{Ours}\end{footnotesize})                                            & \multicolumn{1}{c|}{96.48\begin{tiny}$\pm$0.20\end{tiny}}                                                   & 95.75\begin{tiny}$\pm$0.24\end{tiny}                                                    & \multicolumn{1}{c|}{\textbf{94.83\begin{tiny}$\pm$0.23\end{tiny}}}                                          & \textbf{92.48\begin{tiny}$\pm$0.28\end{tiny}}                                           \\ \hline \hline
\end{tabular}
}
\end{table}

\section{Results}
\label{sec:results}

\subsection{EERs for speaker verification}
\label{subsec:eer_thres}
As discussed in Section (\ref{subsec:fixed_thres}) and shown in Figure (\ref{fig:tsne_plot}), the optimal fixed threshold for different datasets depends on the intrinsic speaker characteristics and data distribution. Table (\ref{tab:eer_thres}) contains the EER for VoxCeleb1 and SITW standard trial pairs and the optimal thresholds corresponding to them. We have also included the EERs for Resnet50 \cite{he2016resnet} and RawNet3 \cite{jung2022pushing} speaker encoders trained on VoxCeleb2 \cite{chung2018voxceleb2}. We can see that the thresholds for SITW are lower compared to VoxCeleb1, which confirms the hypothesis from Figure (\ref{fig:tsne_plot}) that the speaker clusters in SITW are more separable compared to VoxCeleb1. 

\subsection{Imposter identification under domain shift}
\label{subsec:imposter_id_domain_shift}
To show the efficacy of proposed speaker-specific thresholding (SST) and imposter detection network (IDN) and their robustness to domain shift, we compare their performance with fixed thresholding and adaptive score normalization \cite{adaptivenorm2011} on VCTK and VoxCeleb1 (with reverb) in Table (\ref{tab:speaker_id}). We can observe that the proposed approaches outperform both the baselines in almost all the cases. For SST, we always observe that the imposter accuracy is less than the overall accuracy even though we set the speaker-specific threshold as the maximum of inter-speaker similarities in Equation (\ref{eq:speaker_spefific_thres}). This observation, in turn, indicates that the choice of keeping the threshold on the higher end of the region of separation and equal to the maximum of inter-speaker similarities is empirically correct.

Observe that the improvement is not as consistent in the case of VCTK as compared to VoxCeleb1 because the underlying speaker encoder has been trained on a similar dataset, i.e. VoxCeleb2. Although we have used a different set of room impulse responses to augment VoxCeleb1 for testing than those used while training, the model still seems to be robust to it. 

\subsection{Performance on far-field data}

\begin{table}[]
\centering
\caption{Performance of the proposed techniques on the 2022 FFSV Challenge {[}20{]} data for unseen speaker identification. Number of enrolled speakers $M=5$. Refer to Appendix for imposter detection accuracies.}
\label{tab:ffsvc}
\resizebox{0.9\linewidth}{!}{
\begin{tabular}{c|c||c|c|c|c}
\hline \hline
\multirow{2}{*}{\textbf{\begin{tabular}[c]{@{}c@{}}Rec.\\ Device\end{tabular}}} & \multirow{2}{*}{\textbf{\begin{tabular}[c]{@{}c@{}}Mic.\\ Dist.\\ (m)\end{tabular}}} & \textbf{Fixed}                                         & \textbf{\begin{tabular}[c]{@{}c@{}}Score\\ Norm\cite{adaptivenorm2011}\end{tabular}} & \textbf{\begin{tabular}[c]{@{}c@{}}SST \\ (Ours)\end{tabular}} & \textbf{\begin{tabular}[c]{@{}c@{}}IDN\\ (Ours)\end{tabular}} \\ \cline{3-6} 
                                                                              &                                                                                      & \begin{tabular}[c]{@{}c@{}}Overall\\ (\%)\end{tabular} & \begin{tabular}[c]{@{}c@{}}Overall\\ (\%)\end{tabular}        & \begin{tabular}[c]{@{}c@{}}Overall\\ (\%)\end{tabular}         & \begin{tabular}[c]{@{}c@{}}Overall\\ (\%)\end{tabular}        \\ \hline \hline
\multirow{5}{*}{iPhone}                                                       & 0.25                                                                                 & 85.84\begin{tiny}$\pm$0.45\end{tiny}                                                  & \textbf{86.32\begin{tiny}$\pm$0.39\end{tiny}}                                                & 80.92\begin{tiny}$\pm$0.38\end{tiny}                                                          & 82.63\begin{tiny}$\pm$0.31\end{tiny}                                                         \\
                                                                              & 1.0                                                                                  & 81.20\begin{tiny}$\pm$0.42\end{tiny}                                                  & 81.56\begin{tiny}$\pm$0.40\end{tiny}                                                         & 80.51\begin{tiny}$\pm$0.38\end{tiny}                                                          & \textbf{81.80\begin{tiny}$\pm$0.41\end{tiny}}                                                \\
                                                                              & 3.0                                                                                  & 74.32\begin{tiny}$\pm$0.37\end{tiny}                                                  & 75.15\begin{tiny}$\pm$0.35\end{tiny}                                                         & \textbf{77.71\begin{tiny}$\pm$0.33\end{tiny}}                                                 & 76.99\begin{tiny}$\pm$0.38\end{tiny}                                                         \\
                                                                              & 5.0                                                                                  & 77.59\begin{tiny}$\pm$0.46\end{tiny}                                                  & 77.21\begin{tiny}$\pm$0.36\end{tiny}                                                         & 78.70\begin{tiny}$\pm$0.39\end{tiny}                                                          & \textbf{79.12\begin{tiny}$\pm$0.31\end{tiny}}                                                \\
                                                                              & -1.5                                                                                 & 68.93\begin{tiny}$\pm$0.39\end{tiny}                                                  & 69.74\begin{tiny}$\pm$0.42\end{tiny}                                                         & 75.45\begin{tiny}$\pm$0.35\end{tiny}                                                          & \textbf{76.95\begin{tiny}$\pm$0.35\end{tiny}}                                                \\ \hline
\multirow{5}{*}{iPad}                                                         & 0.25                                                                                 & \textbf{83.99\begin{tiny}$\pm$0.36\end{tiny}}                                         & 83.19\begin{tiny}$\pm$0.41\end{tiny}                                                         & 82.47\begin{tiny}$\pm$0.39\end{tiny}                                                          & 81.69\begin{tiny}$\pm$0.35\end{tiny}                                                         \\
                                                                              & 1.0                                                                                  & 81.09\begin{tiny}$\pm$0.32\end{tiny}                                                  & 81.18\begin{tiny}$\pm$0.30\end{tiny}                                                & \textbf{81.89\begin{tiny}$\pm$0.31\end{tiny}}                                                          & 81.61\begin{tiny}$\pm$0.33\end{tiny}                                                         \\
                                                                              & 3.0                                                                                  & 74.30\begin{tiny}$\pm$0.40\end{tiny}                                                  & 75.62\begin{tiny}$\pm$0.36\end{tiny}                                                         & \textbf{78.59\begin{tiny}$\pm$0.39\end{tiny}}                                                 & 77.28\begin{tiny}$\pm$0.32\end{tiny}                                                         \\
                                                                              & 5.0                                                                                  & 76.94\begin{tiny}$\pm$0.33\end{tiny}                                                  & 77.80\begin{tiny}$\pm$0.40\end{tiny}                                                         & 79.51\begin{tiny}$\pm$0.34\end{tiny}                                                          & \textbf{79.65\begin{tiny}$\pm$0.35\end{tiny}}                                                \\
                                                                              & -1.5                                                                                 & 69.66\begin{tiny}$\pm$0.41\end{tiny}                                                  & 70.59\begin{tiny}$\pm$0.38\end{tiny}                                                         & 77.30\begin{tiny}$\pm$0.35\end{tiny}                                                          & \textbf{78.76\begin{tiny}$\pm$0.37\end{tiny}}                                                \\ \hline \hline
\end{tabular}}
\end{table}

To further demonstrate the effectiveness of the proposed technique on domain shift, we report its accuracy on the 2022 FFSVC \cite{FFSVC2022_Eval_Plan} development data in Table (\ref{tab:ffsvc}). We report the performance on different recording devices and recording locations separately to show the effect of domain shift. Distance -1.5 indicate that the recording device is behind the source \cite{FFSVC2022_Eval_Plan}. We have kept the number of enrolled speakers $M=5$. 

We can observe that at a close distance to the microphone (0.25m), baseline techniques perform better than the proposed techniques because proximity to the source means cleaner audio. This can be accounted for the fact that the proposed approach only uses the enrollment samples for computing the thresholds while the fixed threshold is computed using the entire VoxCeleb1 Test Set for all the experiments. As soon as the microphone gets far away from the source and the shift in the domain is more prominent, we see that the proposed techniques perform far better than fixed thresholding. We observe a performance boost of up to 10\% for the proposed techniques for both the recording devices. 

\subsection{Ablation Study}

\begin{table}[]
\centering
\caption{Performance of different versions of the proposed approach for unseen speaker identification on VoxCeleb1 \cite{nagrani2017voxceleb1} (with reverb). RelNet: Relation Network, E2E: End-to-End.}
\label{tab:ablation}
\resizebox{\linewidth}{!}{
\begin{tabular}{c|c||cc|cc}
\hline\hline
\multirow{2}{*}{\textbf{Method}} & \multirow{2}{*}{\textbf{Backend}} & \multicolumn{2}{c|}{\textbf{M=5}}                                                                                                     & \multicolumn{2}{c}{\textbf{M=10}}                                                                                                     \\ \cline{3-6} 
                                 &                                   & \multicolumn{1}{c|}{\begin{tabular}[c]{@{}c@{}}Overall\\ (\%)\end{tabular}} & \begin{tabular}[c]{@{}c@{}}Imposter\\ (\%)\end{tabular} & \multicolumn{1}{c|}{\begin{tabular}[c]{@{}c@{}}Overall\\ (\%)\end{tabular}} & \begin{tabular}[c]{@{}c@{}}Imposter\\ (\%)\end{tabular} \\ \hline\hline
\multirow{2}{*}{Fixed}           & Cosine                            & \multicolumn{1}{c|}{94.64\begin{tiny}$\pm$0.27\end{tiny}}                                                  & 89.52\begin{tiny}$\pm$0.23\end{tiny}                                                   & \multicolumn{1}{c|}{93.06\begin{tiny}$\pm$0.31\end{tiny}}                                                  & 86.47\begin{tiny}$\pm$0.21\end{tiny}                                                   \\
                                 & RelNet                            & \multicolumn{1}{c|}{95.54\begin{tiny}$\pm$0.30\end{tiny}}                                                  & 93.98\begin{tiny}$\pm$0.25\end{tiny}                                                   & \multicolumn{1}{c|}{93.85\begin{tiny}$\pm$0.28\end{tiny}}                                                  & 90.21\begin{tiny}$\pm$0.29\end{tiny}                                                   \\ \hline
\multirow{3}{*}{IDN(\begin{footnotesize}\textbf{Ours}\end{footnotesize})}             & Cosine                            & \multicolumn{1}{c|}{95.86\begin{tiny}$\pm$0.24\end{tiny}}                                                  & 93.29\begin{tiny}$\pm$0.27\end{tiny}                                                   & \multicolumn{1}{c|}{94.01\begin{tiny}$\pm$0.22\end{tiny}}                                                  & 91.07\begin{tiny}$\pm$0.20\end{tiny}                                                   \\
                                 & RelNet \begin{footnotesize}(frozen)\end{footnotesize}                     & \multicolumn{1}{c|}{96.91\begin{tiny}$\pm$0.28\end{tiny}}                                                  & 95.08\begin{tiny}$\pm$0.24\end{tiny}                                                   & \multicolumn{1}{c|}{94.32\begin{tiny}$\pm$0.23\end{tiny}}                                                  & 91.99\begin{tiny}$\pm$0.25\end{tiny}                                                   \\
                                 & RelNet \begin{footnotesize}(E2E)\end{footnotesize}                        & \multicolumn{1}{c|}{\textbf{97.94\begin{tiny}$\pm$0.25\end{tiny}}}                                         & \textbf{96.82\begin{tiny}$\pm$0.29\end{tiny}}                                          & \multicolumn{1}{c|}{\textbf{94.83\begin{tiny}$\pm$0.27\end{tiny}}}                                         & \textbf{92.48\begin{tiny}$\pm$0.20\end{tiny}}                                          \\ \hline\hline
\end{tabular}}
\end{table}

In Table (\ref{tab:ablation}), we show the importance of each of the components of the proposed imposter detection network (IDN) on the VoxCeleb1 \cite{nagrani2017voxceleb1} (with reverb) dataset. Even without a relation network back-end, IDN outperforms the fixed thresholding baselines. We can also observe the performance boost when the IDN is trained end-to-end with relation network compared to when the relation network and speaker encoder are frozen ($4^{th}$ row of Table (\ref{tab:ablation})). This shows that the end-to-end training allows more flexibility to the speaker encoder to be tuned with the IDN.

\section{Conclusions}
\label{sec:conclusion}

In this paper, we  highlighted the problem with using fixed thresholds for imposter detection in unseen speaker identification and proposed a speaker-specific thresholding technique to tackle that. We also introduced a meta-learning framework for end-to-end imposter detection which leverages the enrollment utterances to detect imposters. Furthermore, we show the robustness of the proposed techniques on VCTK, VoxCeleb1 and FFSVC 2022 datasets, and perform ablation study to show the importance of different components.

\bibliographystyle{IEEEbib}
\bibliography{strings,refs}

\end{document}